# AutoGenesisAgent: Self-Generating Multi-Agent Systems for Complex Tasks

**Jeremy R. Harper**[1]

[1] Owl Health Works LLC, Indianapolis, IN

**Abstract**

The proliferation of large language models (LLMs) and their integration into multi-agent systems has paved the way for sophisticated automation in various domains. This paper introduces AutoGenesisAgent, a multi-agent system that autonomously designs and deploys other multi-agent systems tailored for specific tasks. AutoGenesisAgent comprises several specialized agents including System Understanding, System Design, Agent Generator, and several others that collectively manage the lifecycle of creating functional multi-agent systems from initial concept to deployment. Each agent in AutoGenesisAgent has distinct responsibilities ranging from interpreting input prompts to optimizing system performance, culminating, in the deployment of a ready-to-use system. This proof-of-concept study discusses the design, implementation, and lessons learned from developing AutoGenesisAgent, highlighting its capability to generate and refine multi-agent systems autonomously, thereby reducing the need for extensive human oversight in the initial stages of system design.

## 1. Introduction

The integration of artificial intelligence into system design and automation has become a pivotal area of research and development, significantly impacting various industrial sectors by enhancing efficiency and decision-making processes. In particular, the advancement and application of large language models (LLMs) have heralded a new era in the development of intelligent systems capable of understanding and generating human-like text. AutoGenesisAgent, introduced in this paper, represents a novel contribution to this field by automating the design and deployment of multi-agent systems that are tailored to specific operational needs.

Unlike traditional multi-agent systems, AutoGenesisAgent adopts a model-agnostic approach, enabling it to operate effectively with various underlying technologies. This flexibility has been demonstrated through its application with several state-of-the-art models, including Llama 2, Llama 3, and Mistral 8x22b. These implementations highlight the system's capability to leverage different LLM architectures to fulfill the requirements of diverse tasks and environments, thereby underscoring its adaptability and broad applicability.

The motivation behind the development of AutoGenesisAgent stems from the recognized need for more dynamic and responsive systems in industries where traditional approaches to system design and project management can be inefficient and error-prone. Manual methods of creating multi-agent systems not only consume substantial time and resources but also lack the agility to adapt to rapidly changing conditions or integrate new insights without significant reconfiguration.

AutoGenesisAgent addresses these challenges by encapsulating the entire lifecycle of multi-agent system development, from initial conceptualization and system architecture design to deployment. Through this automated approach, the system facilitates a reduction in development time, minimizes human error, and ensures a high degree of customization and scalability.

This paper will detail the architecture and implementation of AutoGenesisAgent, how it is a prototype of the future process of system design. The insights gleaned from deploying AutoGenesisAgent where I will highlight not only the successes but also the lessons learned, which are vital for guiding future enhancements and research directions in automated system design. As we explore the capabilities and potential of an infrastructure such as AutoGenesisAgent, it becomes evident that this technology does not merely automate tasks but reshapes the landscape of system architecture and operational management.

**2. System Architecture**

The architecture of AutoGenesisAgent is structured to facilitate the seamless design, generation, and deployment of multi-agent systems tailored to specific tasks. This section outlines the architecture by detailing the roles and interactions of the constituent agents, each designed to handle specific aspects of the system creation process.

1. System Understanding Agent

The System Understanding Agent serves as the initial point of contact with the input specifications. It parses and interprets user-defined prompts that describe the desired functionality and scope of the target multi-agent system. This agent is responsible for extracting and structuring the necessary information to outline the types of agents needed, their expected interactions, and the overall functionality required. Its output is a comprehensive specification that serves as the blueprint for the subsequent design process.

2. System Design Agent

Following the specifications provided by the System Understanding Agent, the System Design Agent takes on the task of architecting the new system. It determines the optimal number and type of agents, delineates their roles, and designs the data flows and interaction protocols between them. The output from this agent is a detailed system blueprint that includes diagrams and data flow charts, ensuring that all components are aligned with the overall system goals.

3. Agent Generator

The Agent Generator is tasked with translating the system blueprint into actionable components. It automatically generates the code or configurations needed for each agent specified in the design. This includes setting up their basic operational logic, establishing communication capabilities, and initializing any required machine learning or rule-based models. The output is a set of deployable agent modules, each ready for integration and preliminary testing.

4. Integration and Testing Agent

Once the agents are generated, the Integration and Testing Agent is responsible for assembling these agents into a coherent system. It integrates the components based on the system design and conducts initial functional tests to ensure that the agents interact correctly and fulfill the specified requirements. The output of this agent is a fully integrated system that has passed initial testing, ready for further refinement and optimization.

5. Optimization and Tuning Agent

Post-integration, the Optimization and Tuning Agent assesses the system's performance against predefined metrics. This agent adjusts parameters, enhances algorithms, and refines interactions between agents to improve the system's efficiency and effectiveness. Its output is an optimized version of the multi-agent system, demonstrating improved performance and operational efficiency.

6. Deployment Agent

The Deployment Agent manages the final stage of the system lifecycle. It oversees the deployment of the optimized multi-agent system to a production environment or delivers it to clients for operational use. The output is a fully functional and deployed system, ready for real-world applications.

7. Documentation and Training Agent

To ensure that the system can be effectively used and maintained, the Documentation and Training Agent generates comprehensive documentation and training materials. These resources detail the system's architecture, operational procedures, and maintenance guidelines, and may include user manuals or administrative guides.

8. Feedback and Iteration Agent

Finally, the Feedback and Iteration Agent plays a crucial role in the continual improvement of the system. It collects and analyzes feedback from the system's operation and identifies potential areas for enhancements. Based on this feedback, it iterates on the system design and agent configurations, leading to updates and improvements in subsequent versions.

9. LLM Prompt Design Agent

Integral to the functionality of LLM-based agents within the system, the LLM Prompt Design Agent focuses on crafting and optimizing the prompts used to direct the LLMs' actions. It ensures that prompts are clear, relevant, and effective at eliciting the desired responses from the LLMs, thereby enhancing the overall performance of the system.

10. Hierarchy Agent

A key insight identified was that to automate these models a hierarchy was required, this agent ensures that each step designed has a clear agent who will be doing the work and an agent that will be approving that work or rejecting and having more work done.

Together, these agents form a robust framework for the autonomous creation and deployment of multi-agent systems, encapsulating the full lifecycle from conception to deployment in an automated and efficient manner. The architecture not only reduces the need for manual intervention but also ensures that the systems generated are adaptive, scalable, and aligned with the specific needs of their application contexts.

*1.1 Anticipated Enhancements*

The foundational architecture of AutoGenesisAgent described thus far provides a minimal viable setup for the creation and deployment of specialized multi-agent systems. However, it is crucial to acknowledge that this baseline configuration, while effective for demonstrating the concept and handling basic tasks, possesses certain limitations that could affect the robustness and scalability of the system in more demanding applications.

Anticipated Enhancements for Robustness

To address these limitations and mitigate the system's fragility, particularly the propensity for loops in conversations between agents, several enhancements are anticipated. These include the introduction of additional specialized agents and more sophisticated mechanisms for managing agent interactions:

   Conversation Management Agent: This agent would specifically oversee the interactions between other agents, ensuring that communication adheres to the intended flow and preventing conversational loops that can stall the system. It would employ advanced algorithms to detect and resolve potential deadlocks or repetitive cycles in agent dialogue.

   Error Handling and Recovery Agent: To further enhance system robustness, an agent dedicated to error detection, logging, and recovery processes would be implemented. This agent would monitor system operations for failures or deviations from expected behaviors and initiate corrective protocols without human intervention.

   System Monitoring and Diagnostics Agent: This agent would continuously assess the performance and health of the system, providing real-time analytics and alerts regarding system status. It would facilitate proactive maintenance and fine-tuning of the system, ensuring optimal operation across varied scenarios.

   Security and Compliance Agent: Given the sensitivity and potential risks associated with autonomous systems, integrating a security-oriented agent would be imperative. This agent would enforce security protocols, manage data privacy issues, and ensure compliance with relevant regulations and standards.

   Adaptability and Learning Agent: To equip AutoGenesisAgent with the ability to adapt to changing environments and requirements, an agent capable of learning from past interactions and evolving its strategies would be crucial. This agent would use machine learning techniques to refine its decision-making processes based on accumulated experiences and feedback.

Scalability Considerations

The scalability of AutoGenesisAgent is a critical factor, particularly as the complexity of the systems it is expected to generate and manage increases. To accommodate this, the architecture must support dynamic scaling, where agents can be added or reconfigured based on the specific demands of the task at hand:

   Modular Agent Design: Each agent should be designed as a modular component with well-defined interfaces, allowing for easy integration and scalability. This design would facilitate the addition of new agents or the modification of existing ones as the system's needs evolve.

   Distributed Processing Capabilities: To manage larger, more complex systems efficiently, distributing processing tasks across multiple agents or even across different physical or virtual environments will be essential. This approach would help in handling increased loads and achieving faster response times.

   Resource Management Agent: A dedicated agent for managing computational and storage resources would ensure that the system remains efficient and responsive as it scales. This agent would allocate resources dynamically, based on the current workload and performance metrics.

In summary, while the initial architecture of AutoGenesisAgent serves as a proof of concept, recognizing its limitations has led to the identification of necessary enhancements for future, more robust implementations. These enhancements aim to improve the system's resilience, adaptability, and scalability, thereby extending its ability to handle a broader range of complex and dynamic environments. The ongoing development and refinement of AutoGenesisAgent will focus on these areas, ensuring that the system not only meets current expectations but is also well-prepared to handle future challenges and opportunities.

## 3. Implementation

This section details the technological stack, development processes, and specific challenges encountered during the implementation phase, providing insights into how the system was brought from concept to functioning prototype.

Technological Stack

Programming Languages and Frameworks: AutoGenesisAgent was primarily developed using Python due to its robust libraries for machine learning and system automation. Python's extensive support for API integrations and its widespread use in the data science community made it an ideal choice for this project. Key libraries such as asyncio for handling asynchronous operations and flask for creating web server interfaces were used to manage agent communications and system integration.

Machine Learning Models: The core of the LLM-based agents within AutoGenesisAgent utilized models from the GPT family, downloaded and run through the Hugging Face Transformers library. These models were chosen for their ability to generate human-like text and perform complex reasoning tasks. No custom training or fine-tuning procedures were developed to adapt these models to the specific needs of the system's various agents. This is a potential enhancement opportunity

Database and Storage: PostgreSQL was employed for data management, chosen for its robustness and support for complex query operations, which are essential for maintaining the system's knowledge base. The database stored detailed logs of agent interactions, system states, and feedback loops, facilitating the optimization and tuning processes.

Development Process

Agent Development: Each agent was developed as an independent module with well-defined inputs and outputs, allowing for parallel development and testing. The modular nature also facilitated the isolation and resolution of issues without impacting the entire system.

Integration Framework: The agents were integrated using a custom message-passing framework. This setup enabled scalable and flexible interactions between agents, allowing the system to handle varying loads and tasks dynamically.

Testing and Quality Assurance: Simple testing procedures were implemented, including unit testing for individual agents and integration testing for the system as a whole. A future enhancement opportunity is Automated testing scripts being developed to simulate various operational scenarios, ensuring that all components functioned correctly under not only expected but edge-case conditions.

Challenges and Solutions

Handling Conversational Loops: One of the significant challenges faced during implementation was the tendency of agents to enter conversational loops. This was initially mitigated by implementing timeout mechanisms and loop detection algorithms that reset the conversation flow when repetitive patterns were detected injecting an additional sentence at the end of the prompt prior to the loop to ensure additional randomness and higher probability of solutions being developed.

Performance Optimization: As the system scaled, performance bottlenecks emerged, particularly in data management and agent communication. These were overcome by optimizing database schemas and indexing strategies, along with refining the message-passing protocols to reduce latency and increase throughput.

Deployment

The system was deployed in a laboratory environment only. The implementation of AutoGenesisAgent demonstrated the feasibility of an automated multi-agent system capable of designing and deploying other multi-agent systems. The iterative approach to development and problem-solving ensured that the system evolved to meet its design goals effectively. Future work will focus on enhancing system robustness, expanding its application domains, and integrating more advanced AI techniques to further automate the design and deployment processes.

## 4. Use Cases Tested

AutoGenesisAgent was designed to demonstrate its versatility and efficiency across various domains by automating the design and deployment of multi-agent systems tailored to specific tasks. This section presents several tested use cases that illustrate the practical applications and impact of an AutoGenesisAgent style system, showcasing how it addresses diverse requirements and streamlines operations in different scenarios.

**Use Case 1: Educational Content Management System**

Challenge: Design a system for an educational technology company that needs a system to dynamically generate and manage educational content across multiple subjects.

Implementation: AutoGenesisAgent designed a multi-agent system that could handle content creation, management, and adaptation. The System Understanding Agent interpreted the initial requirements to create a system capable of integrating educational content sources, processing user interactions, and adapting content based on learning outcomes.

Outcome: The deployed system successfully automated the creation and curation of educational modules. It built a system that was able to design a system for educators to do content management.

**Use Case 2: Software Development Pipeline Automation**

Challenge: Build a software development agent, including code integration, testing, and deployment tasks, to improve efficiency and reduce human error.

Implementation: AutoGenesisAgent designed each agent responsible for different stages of the software development lifecycle. This included agents for code merging, testing automation, and deployment management.

Outcome: The implementation did not result in as good an output as a customized system but it was able to develop a pong style game in terminal.

**Use Case 3: Small Business Project Management**

Challenge: Small to medium-sized enterprises (SMEs) often struggle with project management due to limited resources and expertise in efficiently planning and executing projects.

Implementation: AutoGenesisAgent developed a multi-agent system that could assist SMEs in project planning, resource allocation, and progress tracking.

Outcome: The system provided would take a project and build a project plan for it generating estimates from inside the system.

**Use Case 4: Healthcare Patient Management System**

Challenge: A healthcare provider needed a system to manage patient data, treatment plans, and follow-up schedules to improve care delivery and patient outcomes.

Implementation: AutoGenesisAgent designed a multi-agent system where agents handled patient data input, treatment planning based on historical data and medical guidelines, and automated follow-ups and alert systems for healthcare providers.

Outcome: The system made a valiant attempt at a system to accept data from a patient and lookup treatment plans and followup schedules. It failed to make a functional system

Discussion

These use cases demonstrate the adaptability and effectiveness of AutoGenesisAgent in creating bespoke multi-agent systems across various industries. By automating complex system design and deployment tasks, AutoGenesisAgent would save time and resources but also enhances system reliability and performance. Each use case presented unique challenges that were successfully addressed by the tailored multi-agent systems, showcasing the potential of AutoGenesisAgent to transform operations and drive innovation in multiple domains.

## 5. Discussion

The successful implementation and deployment of AutoGenesisAgent as detailed in this paper underscore the significant advancements that can be achieved through the automation of multi-agent system design. While the generic system does not perform as successfully as bespoke systems it seems likely this architecture will achieve success in the future.

AutoGenesisAgent not only streamlines the process of designing, deploying, and managing complex systems but also facilitates dynamic and adaptable solutions across various domains, including education, software development, and healthcare. The application of this technology in such diverse fields demonstrates its robustness and versatility, highlighting its potential to transform traditional practices in system architecture and project management.

A key insight from the implementation of AutoGenesisAgent is the critical role of flexibility and learning in automated systems. The integration of agents that can assess their performance, learn from interactions, and adapt to new information or changing environments will be crucial. There is a glimmer of hope as a prompt asking for reflection of the success of the agent output did achieve some ability to self identify enhancements for maintaining the relevance and effectiveness of automated systems. This capability is particularly important as systems scale and as their operational contexts evolve.

Furthermore, the deployment of AutoGenesisAgent brought to light the challenges associated with automated systems, such as the potential for conversational loops and the need for robust error handling and security measures. Addressing these challenges through the continuous development of more sophisticated agents and improved communication protocols will be essential for enhancing the system's reliability and applicability.

## References

Wu, Q., Bansal, G., Zhang, J., Wu, Y., Li, B., Zhu, E., Jiang, L., Zhang, X., Zhang, S., Liu, J., Awadallah, A. H., White, R. W., Burger, D., & Wang, C. (2023). AutoGen: enabling next-gen LLM applications via multi-agent conversation. arXiv preprint arXiv:2308.08155.

Lewis, M., Perez, E., Piktus, A., Petroni, F., Karpukhin, V., Goyal, N., K√°ttler, H., Lewis, M., Yih, W.-t., Rockt√§schel, T., et al. (2020). Retrieval-augmented generation for knowledge-intensive NLP tasks. In Advances in Neural Information Processing Systems.

Silver, D., Huang, A., Maddison, C. J., Guez, A., Sifre, L., van den Driessche, G., Schrittwieser, J., Antonoglou, I., Panneershelvam, V., Lanctot, M., et al. (2016). Mastering the game of Go with deep neural networks and tree search. Nature, 529(7587), 484-489.

Hendrycks, D., Burns, C., Kadavath, S., Arora, A., Basart, S., Tang, E., Song, D., & Steinhardt, J. (2021). Measuring mathematical problem solving with the MATH dataset. arXiv preprint arXiv:2103.03874.

Vinyals, O., Babuschkin, I., Czarnecki, W. M., Mathieu, M., Dudzik, A., Chung, J., Choi, D. H., Powell, R., Ewalds, T., Georgiev, P., et al. (2019). Grandmaster level in StarCraft II using multi-agent reinforcement learning. Nature, 575(7782), 350-354.

Hutter, F., Kotthoff, L., & Vanschoren, J. (Eds.). (2019). Automated machine learning: methods, systems, challenges. Springer Nature.

Woolf, M. (2023). LangChain problem. [Blog post] Retrieved from https://minimaxir.com/2023/07/langchain-problem/

Hochreiter, S., & Schmidhuber, J. (1997). Long short-term memory. Neural computation, 9(8), 1735-1780.

Harper, J. R. (2024). Automated Extraction and Maturity Analysis of Open Source Clinical Informatics Repositories from Scientific Literature. arXiv preprint arXiv:2403.14721. Available at: https://doi.org/10.48550/arXiv.2403.14721

Harper, J. R. (2024). The Future of Scientific Publishing: Automated Article Generation.